\documentclass[preprint]{JHEP3} 

\JHEPspecialurl{http://jhep.sissa.it/JOURNAL/JHEP3.tar.gz}

\usepackage{epsfig,graphicx,multicol,bbm}

\newcommand\fverb{\setbox\pippobox=\hbox\bgroup\verb}
\newcommand\fverbdo{\egroup\medskip\noindent%
			\fbox{\unhbox\pippobox}\ }
\newcommand\fverbit{\egroup\item[\fbox{\unhbox\pippobox}]}
\newbox\pippobox

\title{Elastic and Resonance contributions to moments
of the proton structure function $F_2$}

\author{M.~Osipenko\\
INFN, Sezione di Genova, 16146 Genova, Italy \\
Skobeltsyn Institute of Nuclear Physics, 119992 Moscow, Russia
E-mail: \email{osipenko@ge.infn.it}}

\preprint{\hepph{0307316}}

\abstract{
We discuss the role of nucleon and its excited state poles in the twist expansion
of the nucleon structure function moments. We find that the
nucleon pole contribution was overestimated in previous analyses
by a factor of two. Inclusion of this missing factor makes the duality
appear for all moments at least down to $Q^2=1$ GeV$^2$.
For resonance poles the time reversal invariance
together with unitarity demand at least four of them 
in the forward Compton scattering amplitude. These poles as well as
the elastic pole can be propagated separately through the standard
Operator Product Expansion derivation for DIS.
This part of the amplitude gives a coherent, positive, higher twist contribution,
which can be singled out of the total and compared with the remaining one
given mostly by threshold effects.
A comparison of the estimated resonance contribution to the data on
structure function moments allows to test large-$Q^2$ behavior of
proton-resonance transition form-factors.
}


\begin{document}

\section{\label{sec:intro} Introduction}
Investigations of the nucleon structure with electromagnetic
probes at large momentum transfers allow for a simple treatment
within the parton model. In fact, in this kinematics the inclusive
lepton-nucleon scattering
cross section can be related to the parton momentum distributions
inside the nucleon. Latter do not follow from pQCD
and they are subject to a measurement or Lattice simulations.
Since structure functions themselves cannot be
compared to pQCD predictions one can instead extract quantities
calculable in the theory, moments. The Mellin transform
of structure functions to Euclidean space offers many
simplifications. First of all, it gives an access directly to
the subject of pQCD, the $Q^2$-evolution,
leaving only one undefined parameter to the measurement.
This parameter is the absolute normalization of the moment
at some arbitrary scale. This scale can be chosen in the kinematic
region where approximations involved in calculations are fulfilled.
The second, the calculation of higher order corrections by means of
effective series summation techniques are less involved
in the moment space~\cite{SGR}.
Obviously, there are also difficulties: nucleon
structure functions cannot be measured down to $x=0$.
Therefore, lowest moments always carry
a dose of uncertainty. However, higher
moments are well defined experimentally and they are
subject to precise measurements~\cite{Osipenko_f2p,Osipenko_f2d}.

Scaling down in $Q^2$ we explore a new kinematic region, not yet understood
in terms of pQCD. It is the region of bound partons,
where the interaction with the probe photon likely involves more than
one single parton. Correlated partons can either interact
to collectively produce hadronic final state or
form an excited nucleon state. These multiparton correlations
are responsible for the confinement and they are the subject of
intensive studies~\cite{Osipenko_f2n,Ricco,Ricco2,Rujula,Ji,Bodek,Nikulesku}.
Main goal of these investigations is to estimate the contribution
of multiparton correlations (higher twists) and their
$Q^2$ evolution based on experimental data. The theoretical
basis of these studies~\cite{Ji,Shurak} is still poorly developed.
This creates a large amount of phenomenological approaches
aimed to catch main features of the problem as e.g.
the parton-hadron duality phenomena~\cite{BloGil,Rujula,Ricco,Ricco2,Nikulesku,Isgur}.

In this article we discuss applicability of the Operator
Product Expansion (OPE) to poles of the forward Compton scattering amplitude.
The separation of the pole contribution in the moments
allows to put a lower limit on the higher twist contribution
in the measured moments. Moreover subtraction of the pole
contribution might help to the application of the constituent quark model
picture as proposed in Ref.~\cite{Petronzio}.

\section{\label{sec:fcs} Forward Compton scattering amplitude}
The forward Compton scattering amplitude $T_{\mu\lambda}$ 
of the virtual photon with four-momentum $q$ on the nucleon $N$
is given by~\cite{Roberts}:
\begin{equation}\label{eq:currents}
T_{\mu\lambda}=i \sum_{\sigma} \int d^4 \xi e^{iq\xi}
<N,\sigma| T(J_{\mu}(\xi) J_{\lambda}(0)) |N,\sigma> ~,
\end{equation}
\noindent where the sum is running over polarization degrees
of freedom $\sigma$, $J_{\mu}$ are hadronic currents,
$\xi$ is the light-cone separation between currents
and $T()$ stays for the time-ordered product.
In case of unpolarized scattering off the nucleon
the amplitude can be expressed in terms of two invariant amplitudes $T_1$ and $T_2$:
\begin{equation}
T_{\mu\lambda}=-T_1 \Bigl(g_{\mu\lambda}+\frac{q_{\mu}q_{\lambda}}{Q^2}\Bigr)
+\frac{T_2}{M^2} \Bigl(P_{\mu}+\frac{P\cdot q}{Q^2}q_{\mu}\Bigr)
\Bigl(P_{\lambda}+\frac{P\cdot q}{Q^2}q_{\lambda}\Bigr) ~~~~,
\end{equation}
\noindent where $q_{\mu}=(\nu,\bf{q})$ ($Q^2=-q^2$) and $P_{\mu}$ are the virtual photon
and proton four-momenta, respectively.

Hadronic currents from Eq.~\ref{eq:currents} for particular class of processes
realizing through the production of
an intermediate state particle can be expressed in terms of
invariant elastic and transition form-factors.
The nucleon intermediate state corresponding to the elastic current according to
Ref.~\cite{Weise} can be written as:
\begin{equation}
<N^\prime,\beta| J^{\mu}_{E}(0) |N,\alpha>= \bar{u}_\beta(P^\prime)
\Bigl\{ \gamma^\mu F_1(Q^2) +
i\frac{\sigma^{\mu\nu}q_\nu}{2M} F_2(Q^2) \Bigr\} u_\alpha(P) ~,
\end{equation}
\noindent where $F_1$ and $F_2$ are known Dirac and Pauli form-factors and
$M$ is the proton mass.
The intermediate particle can be also a nucleon resonance.
In this case the hadronic current can be expressed as following~\cite{Thomas}:
\begin{equation}
<N^*_j,\beta|\epsilon^{m}_{\mu} J^{\mu}_{R}(0) |N_{\frac{1}{2}},\alpha>=
\frac{2}{e} \sqrt{M (M^2_{R}-M^2)} A_{m} ~,
\end{equation}
\noindent where $j$ is the resonance spin, $M_{R}$ is the resonance mass,
$A_{m}$ is the resonance helicity amplitude, $\alpha$ and $\beta$
are helicity indices, $m$ indicates the virtual photon helicity state
with respect to the proton helicity defined as in Ref.~\cite{Weise},
and the virtual photon polarization four-vectors are given by:
\begin{equation}
\epsilon^{\pm}_{\mu}=\frac{1}{\sqrt{2}}(0,\pm 1,-i,0), \mbox{       }
\epsilon^0_{\mu}=\frac{1}{Q}(|\vec{q}|,0,0,\nu) ~.
\end{equation}
These two contributions of the amplitude are particular because
represent poles on the Riemann $\nu$-surface.
The rest of the amplitude is an analytic function of $\nu$ describing
the continuum of intermediate states. Therefore, the total amplitude
can be described as a sum of elastic ($T^E$), resonance ($T^R$) and continuum ($T^C$):
\begin{equation}\label{eq:amplitude}
T(\nu,Q^2)=T^E(\nu,Q^2)+T^R(\nu,Q^2)+T^C(\nu,Q^2) ~.
\end{equation}
\noindent
If we plug the complex amplitude from Eq.~\ref{eq:amplitude} into the unitarity
relation $S S^\dagger = 1$ (with $S=1+2iT$) we obtain
\begin{equation}
Im T^E(\nu,Q^2)+Im T^R(\nu,Q^2)+ Im T^C(\nu,Q^2)=
|T^E|^2 + |T^R|^2 + |T^C|^2 + 2 Re [T^R T^{C\dagger}] ~,
\end{equation}
\noindent collecting the same singularities in r.h.s. and l.h.s.
and neglecting the interference effects we obtain the optical
theorem for each individual term separately.

In this article we are interested in the case of $T_2$.
We assume that elastic and resonance amplitudes are simple poles
and can be written in the factorized form:
\begin{equation}
\nu T_2^{E,R} = R^{E,R}(Q^2) P^{E,R}(\nu) ~,
\end{equation}
\noindent where the residue of the elastic pole is given by
\begin{equation}
R^E(Q^2) = \frac{G_E^2(Q^2)+\tau G_M^2(Q^2)}{1+\tau} ~,
\end{equation}
\noindent where Sachs form-factors $G_E$ and $G_M$ are related to
$F_1$ and $F_2$ as following:
\begin{equation}
G_E=F_1-\frac{Q^2}{4M^2}F_2, \mbox{       }
G_M=F_1+F_2 ~.
\end{equation}
While resonance residues can be written as
\begin{equation}
R^R(Q^2) = \frac{1}{2\pi\alpha}
\frac{M_R^2-M^2}{4M}\frac{Q^2}{|\bf{q_R}|^2}
\Bigl[|A_{1/2}|^2+|A_{3/2}|^2+
2\frac{Q^2}{|\bf{q_R}|^2}|S_{1/2}|^2\Bigr] ~,
\end{equation}
\noindent here $\bf{q_R}$ is the virtual photon three-momentum
taken at the resonance pole.

\section{\label{sec:rprs} Pole structure of the Compton amplitude}
The Compton amplitude defined in the first quadrant of the complex $\nu$-plane
can be analytically continued in other quadrants by means of the crossing
symmetry. The analytic continuation can be performed through the following
relation between crossing channels: $T_2(-\nu)=T_2(\nu)$ (time reversal)
and $T_2(\nu^*)=T_2^*(\nu)$ ($\gamma\gamma\rightarrow p \bar{p}$)\footnote{in
the following two sections we will drop Lorenz indices for simplicity}.
The Compton amplitude has a number of simple poles and branch cuts shown in Fig.~\ref{fig:contour}.
All these poles and cuts should obey crossing symmetry relations mentioned above.
The elastic scattering amplitude has two poles situated on the real axis
at $\nu=\pm \nu_E=\pm Q^2/2M$ and it has the following simple structure:
\begin{equation}
P^E(\nu)=\frac{\nu}{\nu_E-\nu}+\frac{\nu}{\nu_E+\nu} ~.
\end{equation}

\FIGURE[t]{
\includegraphics[bb=1cm 1cm 22cm 20cm, scale=0.3]{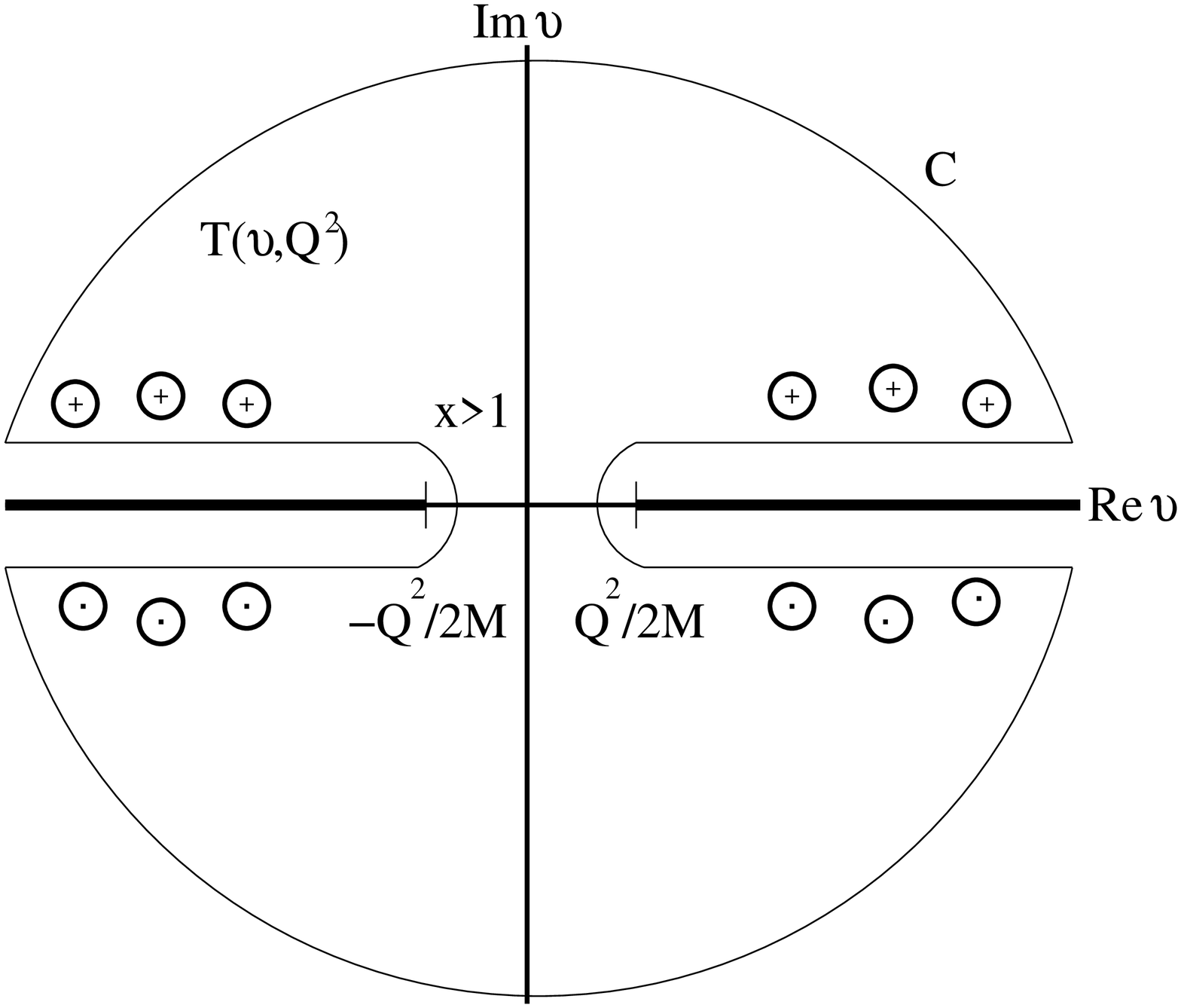}
\caption{\label{fig:contour}
The integration contour $C$ contains usual $\mbox{Re} \mbox{ }\nu$ branch cuts from $\pm Q^2/2M$
up to $\pm\infty$ and poles of the nucleon excited states.
Poles are situated on the unphysical Riemann sheets (poles indicated with dot are located at lower
sheet and those with crosses are on the upper sheet) and do not appear within the contour C.}
}

Nucleon resonances also generate poles in the Compton amplitude.
But these are located on unphysical Riemann sheets. In order to satisfy symmetry
relations each resonance should have at least four poles, two on each
lower and upper unphysical Riemann sheets \footnote{actually poles are present on many
lower and many upper unphysical sheets generated by different decay channels}.
Nevertheless, we limit ourselves to simple poles, which do not carry information
about Riemann sheet of location\footnote{in principle the pole structure of the resonance
amplitude is more complex: poles should ``remember'' the Riemann sheet
which they belong to. This can be parameterized in various forms. One of the
simplest solution is the following:
\begin{equation}\label{eq:n_poles}
P^R(\nu)=-i \frac{\pm \sqrt{\nu_E-\nu_R}}{\sqrt{\nu_E \pm \nu} - \sqrt{\nu_E-\nu_R}} ~,
\end{equation}
\noindent plus two conjugate poles at $\nu_R^\star$,
here $\pm$ signs refer to the principal and secondary square root values
indicating therefore correct Riemann sheet. However, in this study we are interested
in the amplitude at small $\nu<<\nu_E$ where the difference between two
cases is not so important but the simplification of using simple poles
is instead very significant}:
\begin{equation}\label{eq:bw_poles}
P^R(\nu)=
\Biggl|\frac{\nu}{\nu_R-\nu}+\frac{\nu}{\nu_R+\nu}\Biggr|_{+i\varepsilon}
+\Biggl|\frac{\nu}{\nu^*_R-\nu}+\frac{\nu}{\nu^*_R+\nu}\Biggr|_{-i\varepsilon} ~,
\end{equation}
\noindent where $\varepsilon$ is an infinitesimal
positive constant indicating Riemann sheet of influence
and the resonance is located at:
\begin{equation}
\nu_R=\frac{Q^2+(M_R^2-M^2)}{2M}+i\frac{\Gamma_R M_R}{2M} ~,
\end{equation}
\noindent where $\Gamma_R$ is its width.
One can check that Eq.~\ref{eq:bw_poles} yields the familiar Breit-Wigner
parameterization.
\FIGURE[t]{
\includegraphics[bb=3cm 22cm 15cm 27cm, scale=1.0]{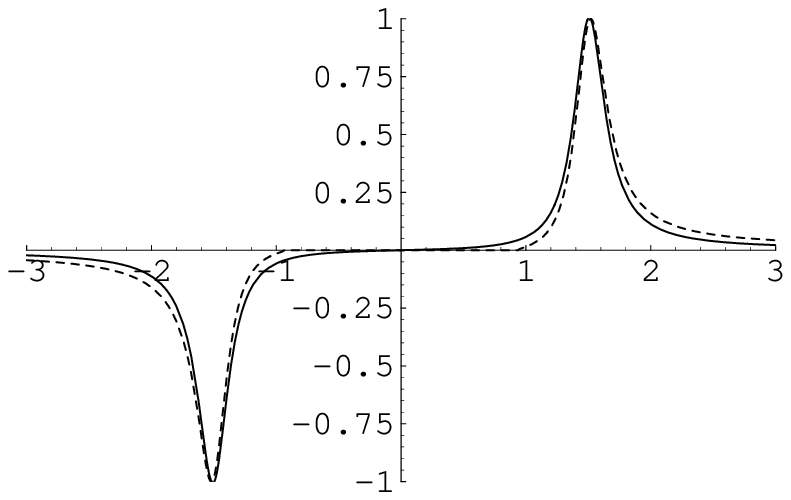}
\caption{\label{fig:pfunc}
The imaginary part of the resonance shape distribution:
the solid line shows the shape of Eq.~\ref{eq:bw_poles} (Breit-Wigner)
and the dashed line represents Eq.~\ref{eq:n_poles}.
Note that the dotted line goes to zero at the threshold.}
}

Unfortunately the simple parameterization from Eq.~\ref{eq:bw_poles} does not
respect the second crossing symmetry property: $T_2(\nu^*)=T_2^*(\nu)$
and the amplitude becomes imaginary on the real axis even in the region
$|\nu|<Q^2/2M$ where no branch cut is present. We can adjust this by brute
force method multiplying the $+i\varepsilon$ term in Eq.~\ref{eq:bw_poles} by:
\begin{equation}\label{eq:bwm_poles}
G=\left\{
\begin{array}{ll}
-1, & |\nu |>Q^2/2M \\
 1, & |\nu |<Q^2/2M ~.
\end{array}
\right.
\end{equation}

This picture is in contrast with Refs.~\cite{Drehsel_oth} where only poles on the lower
unphysical sheet are predicted. However, we find that the absence of poles
on the upper unphysical sheet would generate a disbalance between lower and
upper semi-planes of the Compton amplitude on the physical sheet.
This disbalance would result in an imaginary part
of the discontinuity across the branch cut and therefore would violate the optical theorem:
\begin{equation}
Disc (T^R)=\frac{1}{2i}\Bigl[T^R(\nu+i\varepsilon)-T^R(\nu-i\varepsilon)\Bigr] \nonumber \\
=\frac{1}{2i}T^R(\nu+i\varepsilon) \in \mbox{\bf C} \neq 2\pi W^R(\nu) \in \mbox{\bf R} ~,
\end{equation}
\noindent where $W^R$ is the resonance part of hadronic tensor.

\section{\label{sec:lee} Low Energy Expansion}
The standard OPE derivation for DIS~\cite{Roberts} proceeds through an expansion
of the dispersion integral in the series of $\nu$ in unphysical kinematic domain $|\nu|<Q^2/2M$,
which we call here ``low energy'' regime.
The Cauchy integral over the contour {\bf C} shown in Fig.~\ref{fig:contour}
is given by:
\begin{equation}\label{eq:disp_rel}
T(Q^2,\nu)=\frac{1}{2\pi i}
\int_C \frac{T(Q^2,\nu^{\prime})d\nu^{\prime}}{\nu^{\prime}-\nu} ~.
\end{equation}
\noindent Assuming that the integral over the circle of radius $R\rightarrow\infty$ vanishes
and taking discontinuities across left and right branch cuts,
using also the time reversal invariance $T(Q^2,-\nu)=T(Q^2,\nu)$ we obtain:
\begin{eqnarray}
&& \int_C \frac{T(Q^2,\nu^{\prime})d\nu^{\prime}}{\nu^{\prime}-\nu}=
2 i \int_{\nu_E}^{\infty}\frac{Disc(T(Q^2,\nu^{\prime}))d\nu^{\prime}}{\nu^{\prime}-\nu}
+2 i \int_{-\infty}^{-\nu_E}\frac{Disc(T(Q^2,\nu^{\prime}))d\nu^{\prime}}{\nu^{\prime}-\nu} \nonumber \\
&& =2 i \int_{\nu_E}^{\infty} \Biggl\{ \frac{1}{\nu^{\prime}-\nu}+\frac{1}{\nu^{\prime}+\nu} \Biggr\}
Disc(T(Q^2,\nu^{\prime}))d\nu^{\prime} ~.
\end{eqnarray}
\noindent 
From the other hand the optical theorem states that:
\begin{equation}\label{eq:opteor}
Disc (T(Q^2,\nu))=Im (T(Q^2,\nu+i\varepsilon)) = 2\pi W(Q^2,\nu) ~.
\end{equation}
\noindent
Therefore, substituting the discontinuity $Disc(T(Q^2,\nu))$
by Eq.~\ref{eq:opteor} we can rewrite
the dispersion relation in the following form:
\begin{equation}
T(Q^2,\nu) =
4\int_{Q^2/2M}^{\infty}\frac{d\nu^{\prime}\nu^{\prime}}{\nu^{\prime 2}-\nu^2}
\Biggl\{W(Q^2,\nu^{\prime})\Biggr\} ~.
\end{equation}
Choosing $|\nu|<Q^2/2M$ one expands this into the geometrical series:
\begin{equation}\label{eq:amp_expand_tot}
T(Q^2,\nu) = 4 \sum_{n=0,even}^\infty \tilde{M}_n(Q^2) \Bigl[\frac{1}{x}\Bigr]^n ~,
\end{equation}
\noindent where
\begin{equation}
\tilde{M}_n(Q^2)=\int_0^1 dx^{\prime} x^{\prime n+1}W(Q^2,x^{\prime}) ~.
\end{equation}

In turn, the OPE of the product of two hadronic currents separated by a small light-cone
distance is given by~\cite{Roberts}:
\begin{equation}
T(Q^2,\nu)=\sum_{\tau,n}C_{\tau n}(Q^2,\mu^2)O_n^{\tau}(\mu^2)
\Biggl(\frac{1}{x}\Biggr)^n
\Biggl(\frac{1}{Q^2}\Biggr)^{\tau/2-1} ~,
\end{equation}
\noindent and making the standard correspondence between terms
with equal powers of $1/x$ we obtain usual DIS twist expansion:
\begin{equation}\label{eq:opemod}
\tilde{M}_n(Q^2)=
\frac{1}{4} \sum_{\tau=2}^{\infty} C_{\tau,n}(Q^2,\mu^2)O_n^{\tau}(\mu^2)
\Bigl(\frac{1}{Q^2}\Bigr)^{\tau/2-1} ~,
\end{equation}
\noindent where expansion coefficients $C_{\tau,n}$ can be analytically
calculated in pQCD and local operators $O_n^\tau$ can be obtained from
Lattice simulations.

In order to apply this expression in practice we need to rewrite
Eq.~\ref{eq:opemod} for a measured structure function. For instance,
it can be done for
moments of the proton structure function $F_2=\nu W_2$
extracted from the data in Ref.~\cite{Osipenko_f2p}.
$n$-th moment of the structure function $F_2$ is defined as:
\begin{equation}\label{eq:CWMoment}
M_n(Q^2)=\int_0^1dxx^{n}F_2(x,Q^2) ~,
\end{equation}
\noindent and therefore one obtains the following expansion:
\begin{equation}\label{eq:normope}
M_n(Q^2)=\sum_{\tau=2k}^{\infty}E_{n \tau}(\mu,Q^2)
O_{n \tau}(\mu)\biggl(\frac{\mu^2}{Q^2}\biggr)^{{1 \over 2}(\tau-2)} ~,
\end{equation}
\noindent
where $k=1,2,...,\infty$, $n=2,4...,\infty$\footnote{$n=0$ relation is meaningless
because Regge theory predicts that the structure function behaves as $1/x$
at $x\rightarrow 0$ and therefore $n=0$ integral in Eq.~\ref{eq:CWMoment} would be diverging.},
$\mu$ is a reference scale, $O_{n \tau}(\mu)$ is the reduced matrix element
of the local operators with definite spin $n$ and twist $\tau$ (dimension minus spin),
related to the non-perturbative structure of the target.  $E_{n \tau}(\mu,Q^2)$
is a dimensionless coefficient function describing the small distance behavior,
which can be perturbatively expressed as a power expansion of the running
coupling constant $\alpha_s(Q^2)$.

However, we noticed that also the pole (elastic and resonance) amplitude
contributing to the l.h.s of Eq.~\ref{eq:amp_expand}
can be expanded in the power series of $\nu$ in the region $|\nu|<Q^2/2M$:
\begin{equation}
\nu T_2^E(\nu,Q^2)= 2 R^E(Q^2) \sum_{n=1,odd}^\infty \Bigl[\frac{\nu}{\nu_E}\Bigr]^n
=\frac{2}{x} R^E(Q^2)\sum_{n=0,even}^\infty\Bigl[\frac{1}{x}\Bigr]^n
\end{equation}
\noindent and
\begin{equation}
\nu T_2^R(\nu,Q^2)= 4 R^R(Q^2) \sum_{n=1,odd}^\infty
\Bigl[\frac{\nu}{|\nu_R|}\Bigr]^n \cos{(n\phi_R)}
=\frac{4}{x} |x_R| \sum_{n=0,even}^\infty \Bigl[\frac{|x_R|}{x}\Bigr]^n \cos{((n+1)\phi_R)} ~,
\end{equation}
\noindent where $\phi_R= atan{\frac{\Gamma_R M_R}{Q^2+(M_R^2-M^2)}}$ is the phase of $\nu_R$.

In case of $T_2$ amplitude the moment expansion will give
\begin{equation}\label{eq:amp_expand}
\nu T_2(Q^2,\nu) = \frac{4}{x} \sum_{n=0,even}^\infty M_n(Q^2) \Bigl[\frac{1}{x}\Bigr]^n ~.
\end{equation}
\noindent We can rewrite Eq.~\ref{eq:amp_expand} as following:
\begin{equation}
\nu T_2^C(\nu,Q^2)=\frac{4}{x} \sum_{n=0,even}^\infty M_n(Q^2) \Bigl[\frac{1}{x}\Bigr]^n
-\nu T_2^E(\nu,Q^2)-\nu T_2^R(\nu,Q^2) ~,
\end{equation}
\noindent therefore the continuum part of the Compton amplitude is given by
\begin{equation}
\nu T_2^C(\nu,Q^2)= \frac{4}{x} \sum_{n=0,even}^\infty \Bigl[\frac{1}{x}\Bigr]^n
\Bigl\{ M_n(Q^2)
-\frac{1}{2} R^E(Q^2)
- R^R(Q^2) |x_R|^{n+1} \cos{((n+1)\phi_R)} \Bigr\} ~,
\end{equation}
\noindent or in terms of measured moments\footnote{notice that the definition in Eq.~\ref{eq:CWMoment}
is different from that of the Ref.~\cite{Osipenko_f2p} and we replace here $n$ with $n-2$ for consistency}
it can be rewritten
\begin{equation}\label{eq:res_moms}
M^C_n(Q^2)= M_n(Q^2)-\frac{1}{2} F_2^E(Q^2) -
R_R(Q^2) |x_R|^{n-1} \cos{((n-1)\phi_R)} ~,
\end{equation}
\noindent where the last two terms can be estimated using phenomenological data
on elastic and resonance transition form-factors.

\section{Results}
First of all one can notice that the elastic contribution
comes out to be a factor of two smaller than in all previous
works (see for example Ref.~\cite{Rujula}). Previously, the
dispersion relation method was applied uniformly to both the elastic
and inelastic channels. This led to a confusion in two aspects:
1) the discontinuity across a pole is not a well defined object
from the mathematical point of view;
2) the value of the integral along the semi-circle around the pole
is not vanishing.
One can prove this statement by evaluating the dispersion
relation for the elastic amplitude alone. If the elastic amplitude is given by:
\begin{equation}\label{eq:epole}
T(\nu )=\frac{1}{\nu_E-\nu}+\frac{1}{\nu_E+\nu} ~,
\end{equation}
\noindent then applying the dispersion relation from Eq.~\ref{eq:disp_rel}
using the discontinuity across the pole one finds:
\begin{equation}\label{eq:edisper}
\frac{1}{\nu_E-\nu}+\frac{1}{\nu_E+\nu} \neq 
\frac{1}{\pi}\int_0^\infty \Biggl[ \frac{1}{\nu-\nu^\prime}-
\frac{1}{\nu+\nu^\prime} \Biggr] \times
Im T(\nu^\prime) d\nu^\prime =
2\Biggl [ \frac{1}{\nu_E-\nu} + \frac{1}{\nu_E+\nu} \Biggr] ~.
\end{equation}
\noindent
Therefore, the evaluation of the discontinuity across the elastic pole
in the dispertion relation
leads to the amplitude which is twice larger than the original one.
The comparison of moments obtained in the present article and those from Ref.~\cite{Rujula}
is shown in the Fig.~\ref{fig:mdual} together with various pQCD calculations.

\FIGURE[t]{
\includegraphics[bb=1cm 4cm 20cm 24cm, scale=0.4]{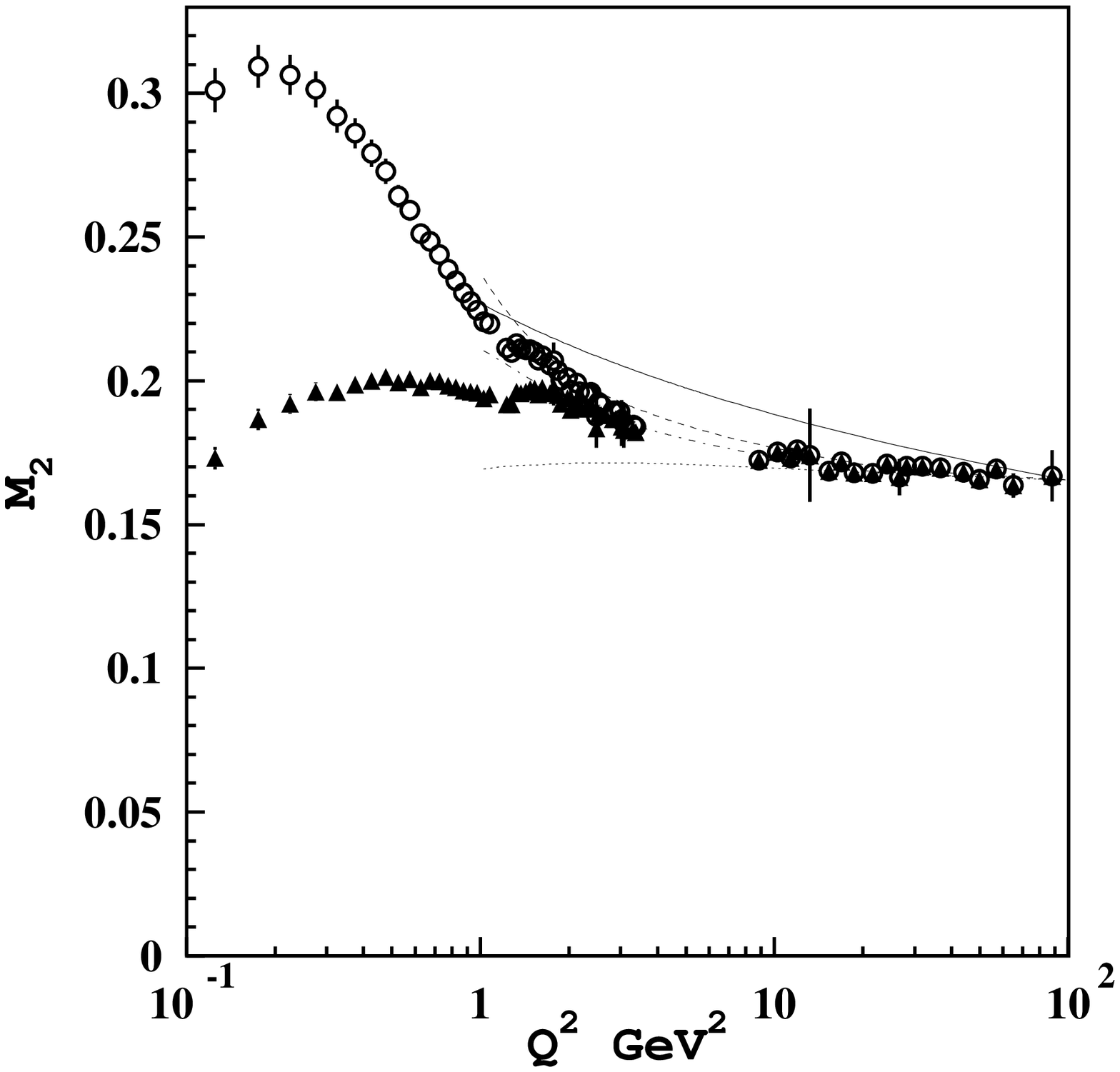}
\includegraphics[bb=1cm 4cm 20cm 24cm, scale=0.4]{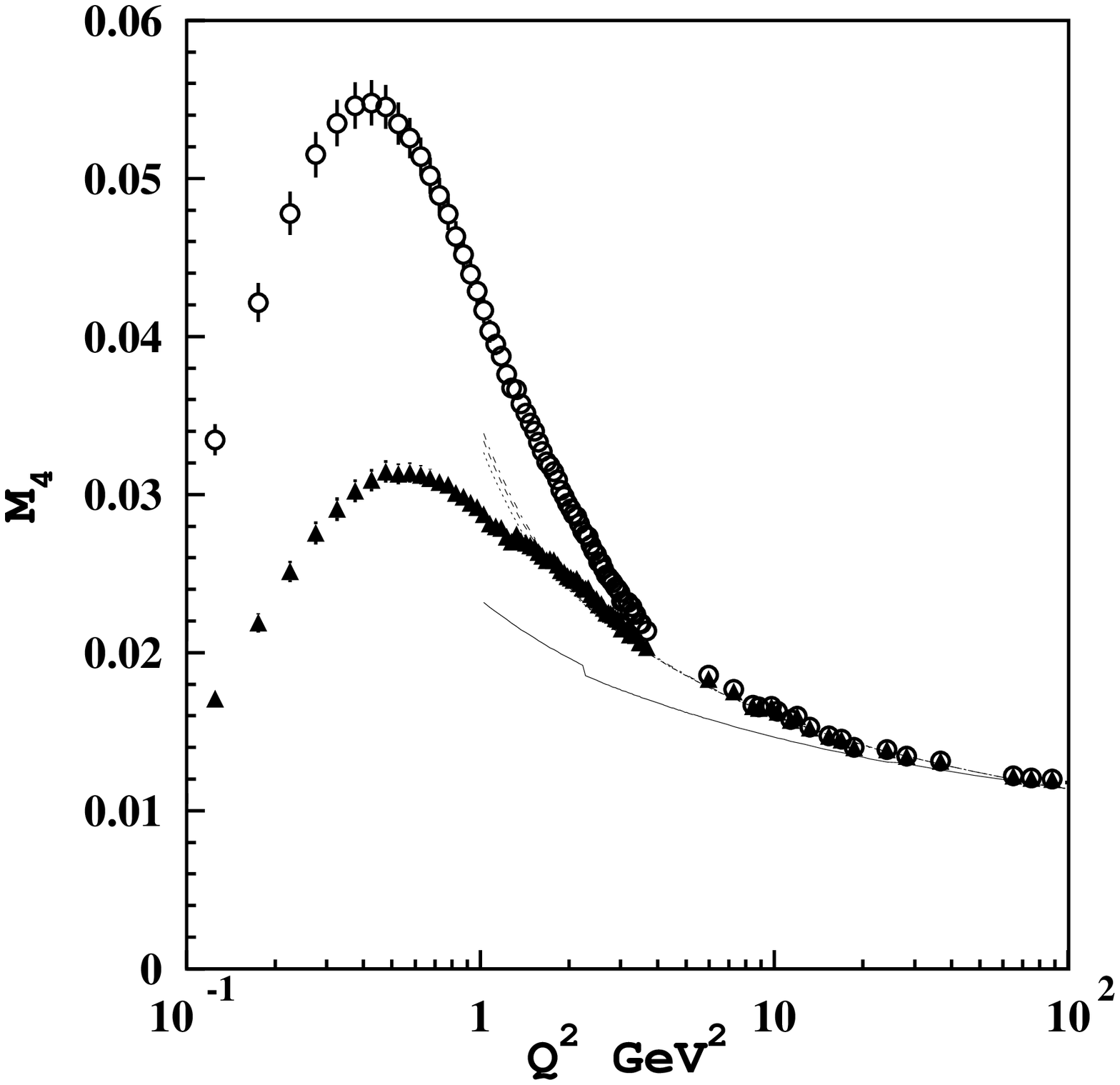}
\includegraphics[bb=1cm 4cm 20cm 24cm, scale=0.4]{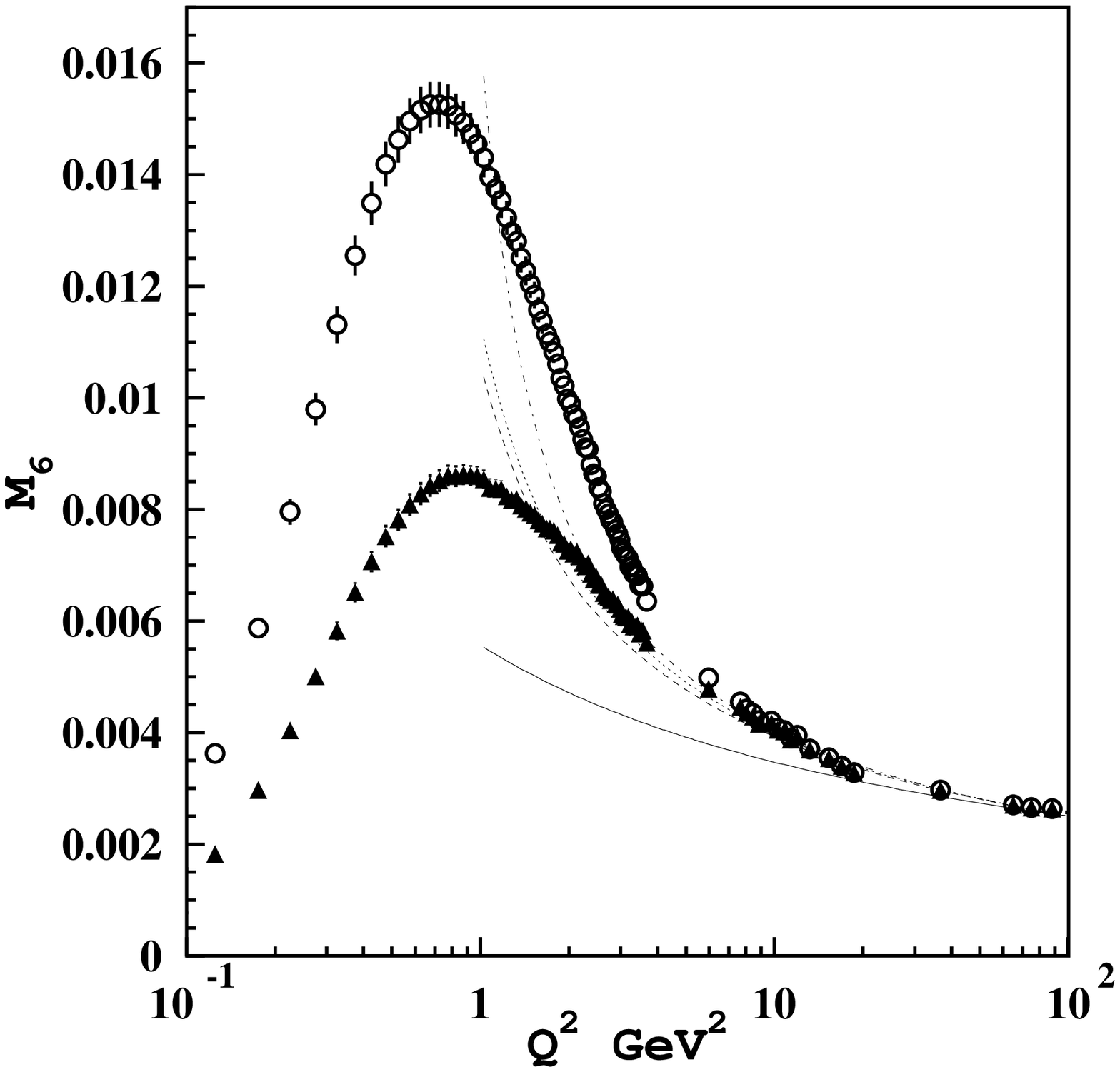}
\includegraphics[bb=1cm 4cm 20cm 24cm, scale=0.4]{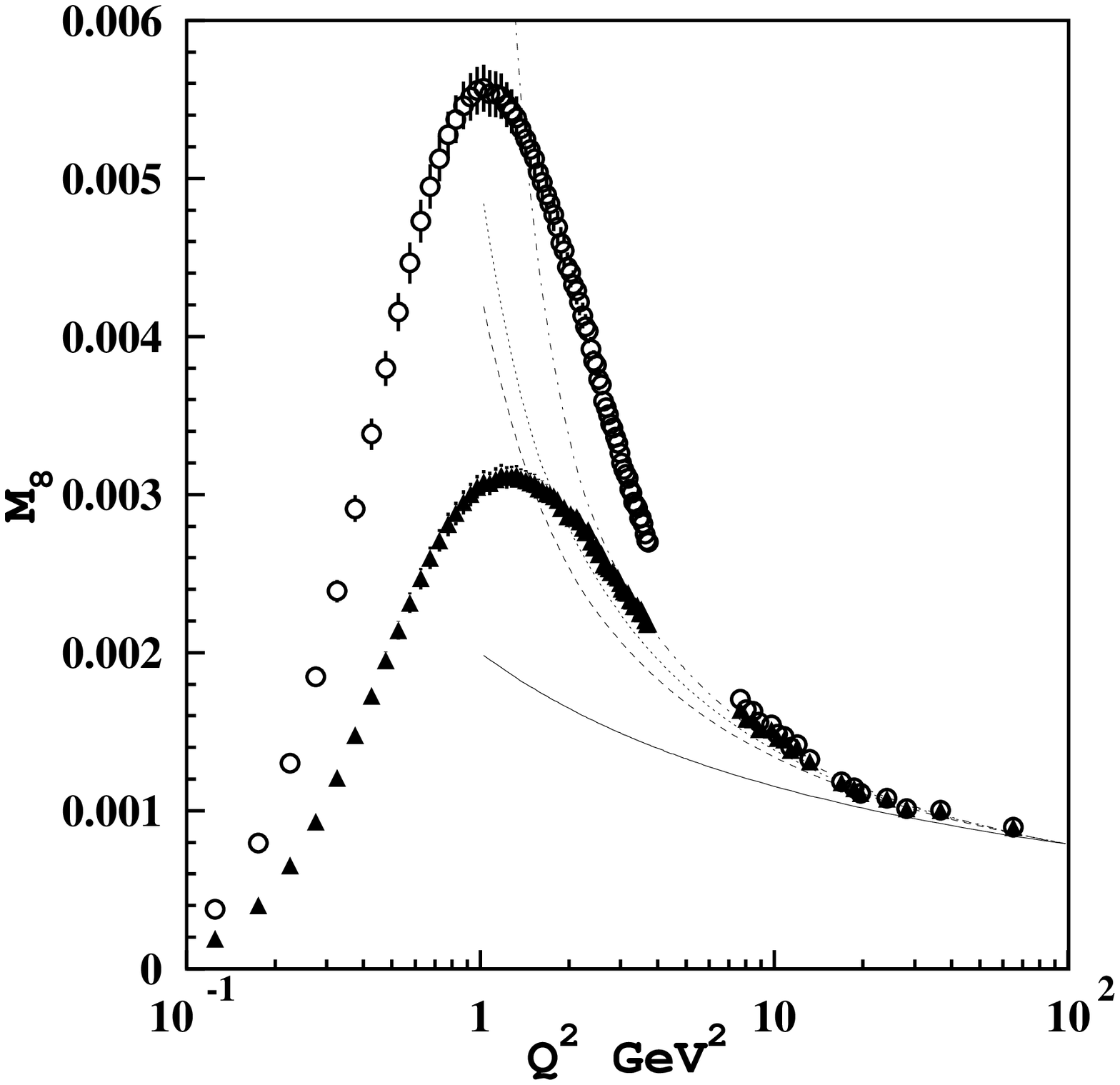}
\caption{\label{fig:mdual} Total Nachtmann moments
of the proton structure function $F_2$ from Ref.~\cite{Osipenko_f2p}
obtained by adding the elastic contribution according to Eq.~\ref{eq:res_moms}
(solid triangles) and ``standard'' formula (empty circles).
The lines represent pQCD calculation fitted to the data at largest $Q^2$:
solid line - LO, dashed line - NLO, dotted line - NNLO,
dot-dhashed - NLO+SGR from Ref.~\cite{SGR}.}
}

Using the relation from Eq.~\ref{eq:res_moms} we estimated the resonance
contribution to the proton structure function moments.
For the sake of simplicity we neglected the longitudinal
couplings $S_{1/2}$ of nucleon resonances.
For this exploratory work this is a good approximation,
provided that those couplings are typically small.
We used phenomenological parameterizations
of the nucleon resonance helicity amplitudes from Ref.~\cite{SQTM}
and Ref.~\cite{Simula_ffs}. We have taken into account 9 (20) well established
resonances from the parameterization~\cite{Simula_ffs} (\cite{SQTM})
whose masses and widths were obtained from Ref.~\cite{PDG}.
In order to compare this to the data extracted in Ref.~\cite{Osipenko_f2p}
we had to convert Conwell-Norton moments into Nachtmann ones.
In the region $M^2/Q^2<1$ this can be done by expanding Nachtmann
moments in a power series of $M^2/Q^2$ as in the Ref.~\cite{Melnitc_tmc}.
Results of the comparison are shown in Figs.~\ref{fig:min}
and Fig.~\ref{fig:rmin}.

\FIGURE[t]{
\includegraphics[bb=1cm 4cm 20cm 24cm, scale=0.4]{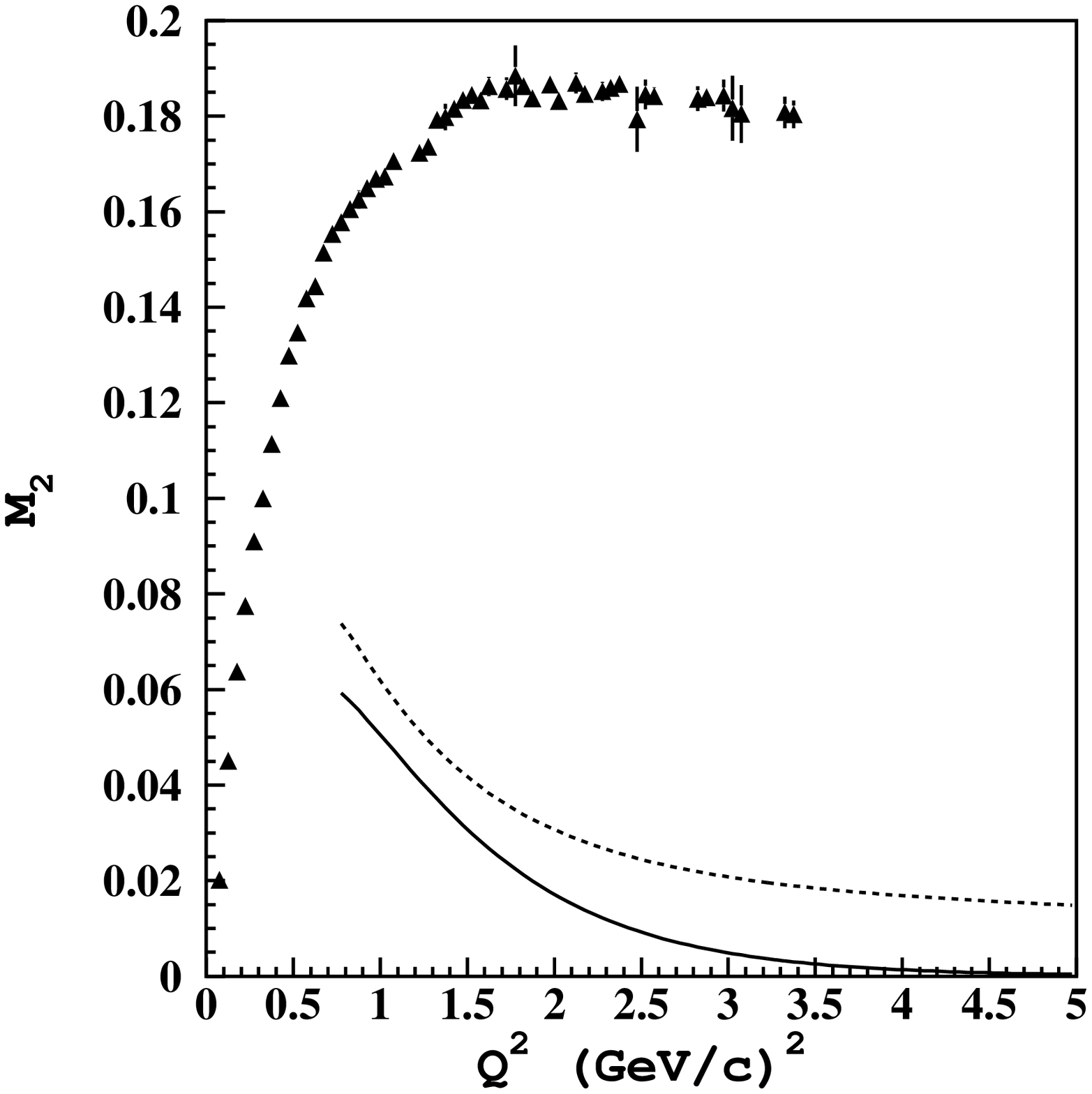}
\includegraphics[bb=1cm 4cm 20cm 24cm, scale=0.4]{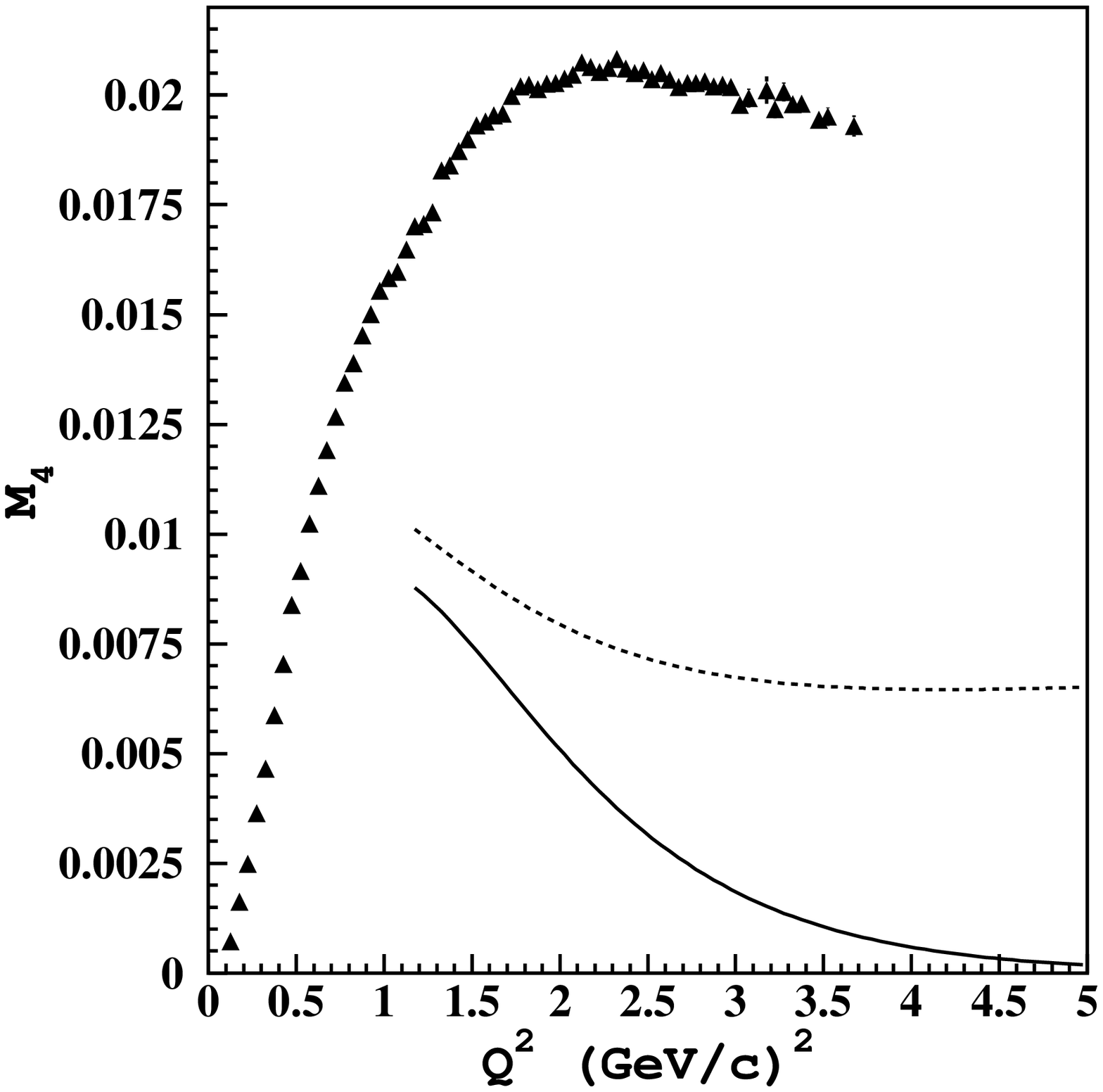}
\includegraphics[bb=1cm 4cm 20cm 24cm, scale=0.4]{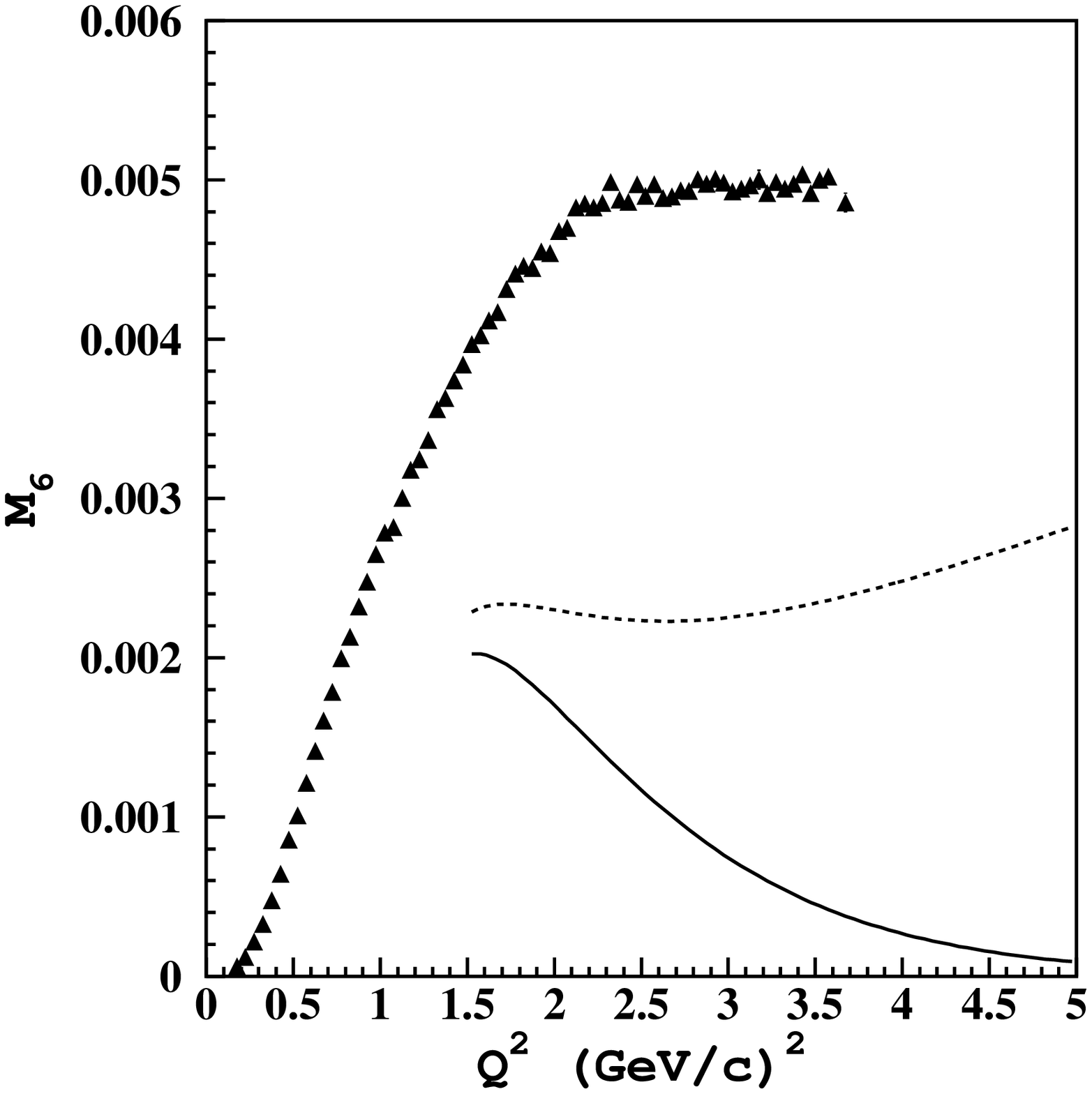}
\includegraphics[bb=1cm 4cm 20cm 24cm, scale=0.4]{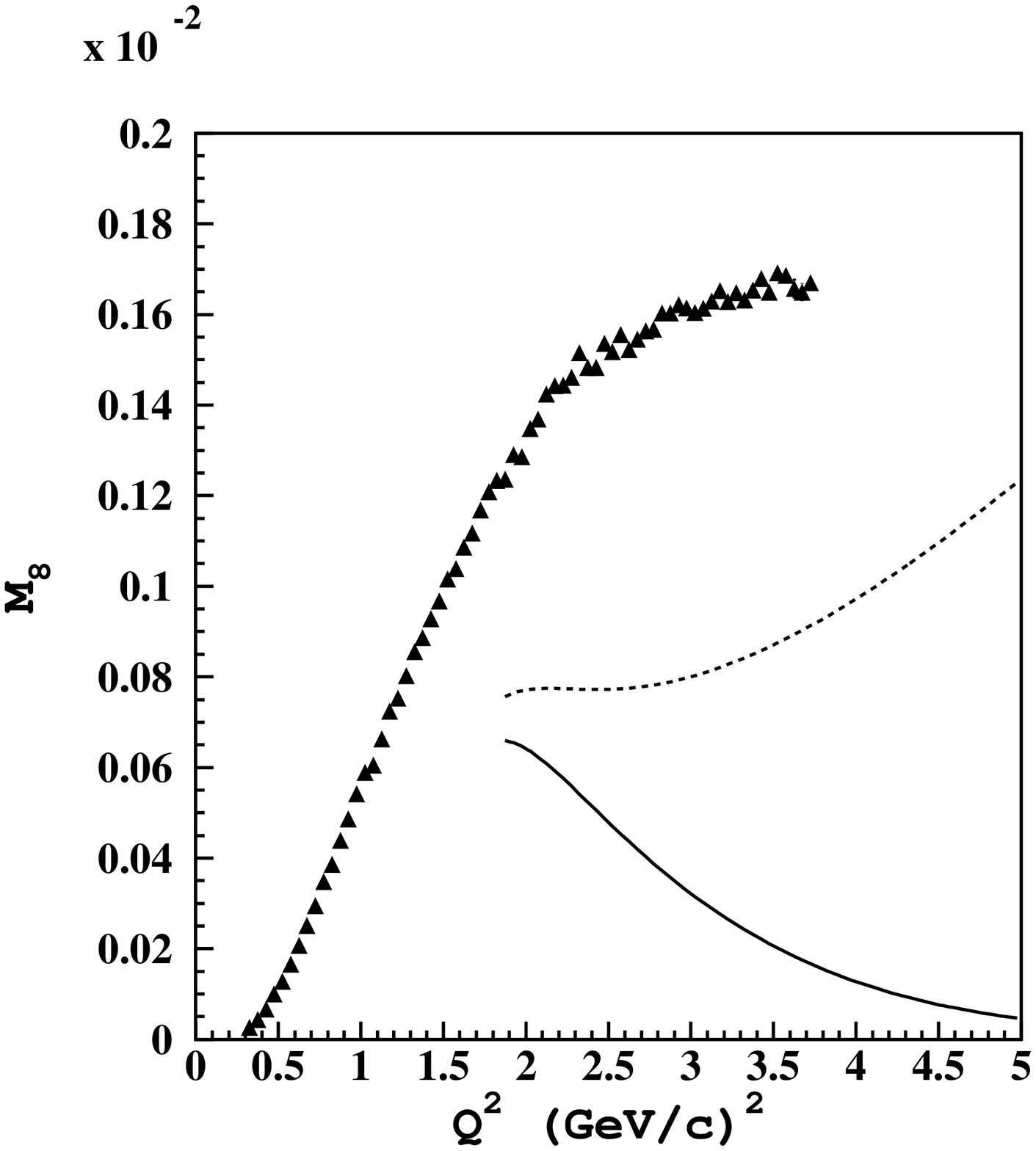}
\caption{\label{fig:min} Inelastic Nachtmann moments
of the proton structure function $F_2$ from Ref.~\cite{Osipenko_f2p}:
solid line represents the resonance contribution with form-factors of Ref.~\cite{SQTM};
dashed line shows the resonance contribution using form-factors from Ref.~\cite{Simula_ffs}.}
}

\FIGURE[t]{
\includegraphics[bb=1cm 4cm 20cm 24cm, scale=0.4]{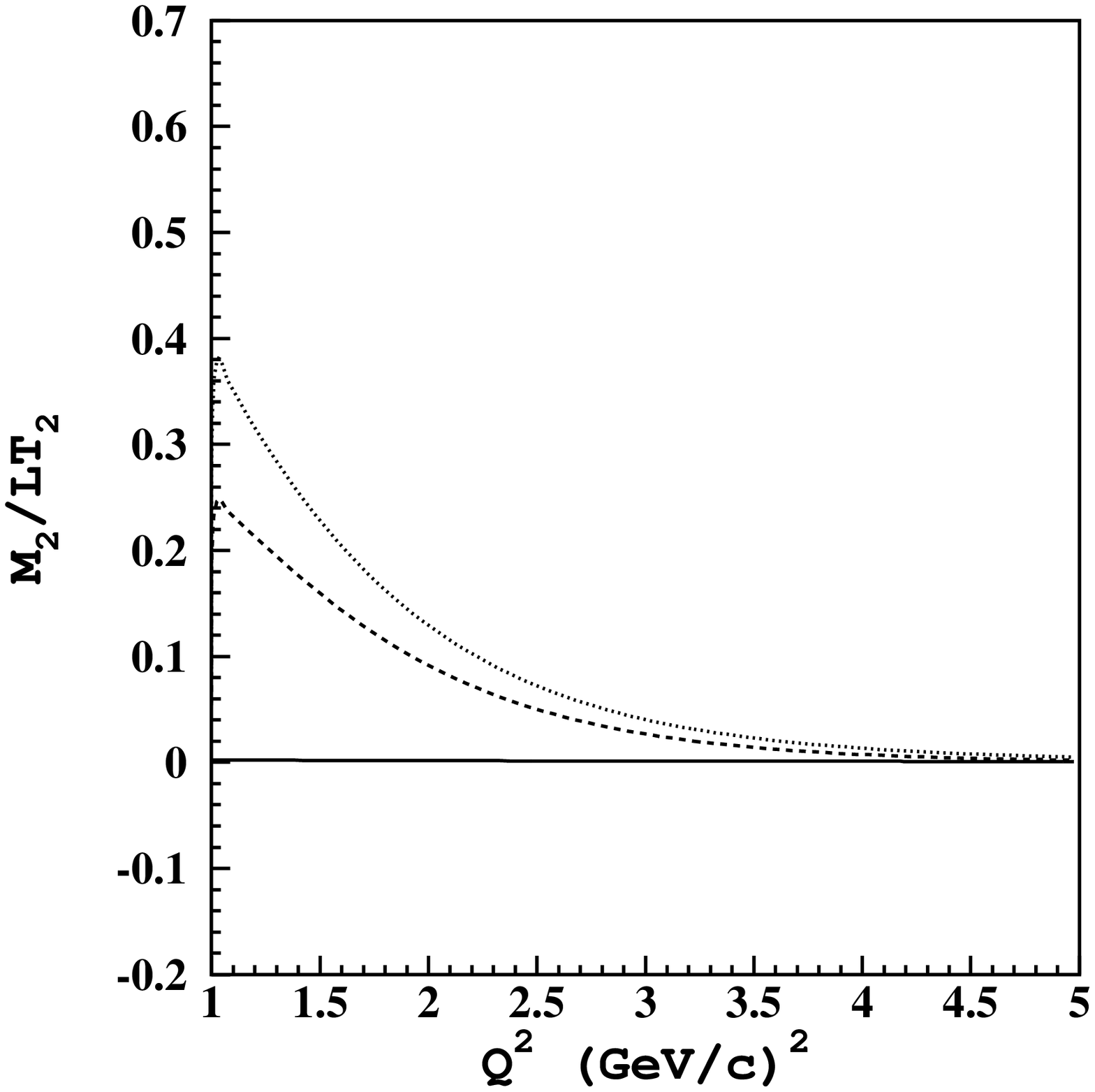}
\includegraphics[bb=1cm 4cm 20cm 24cm, scale=0.4]{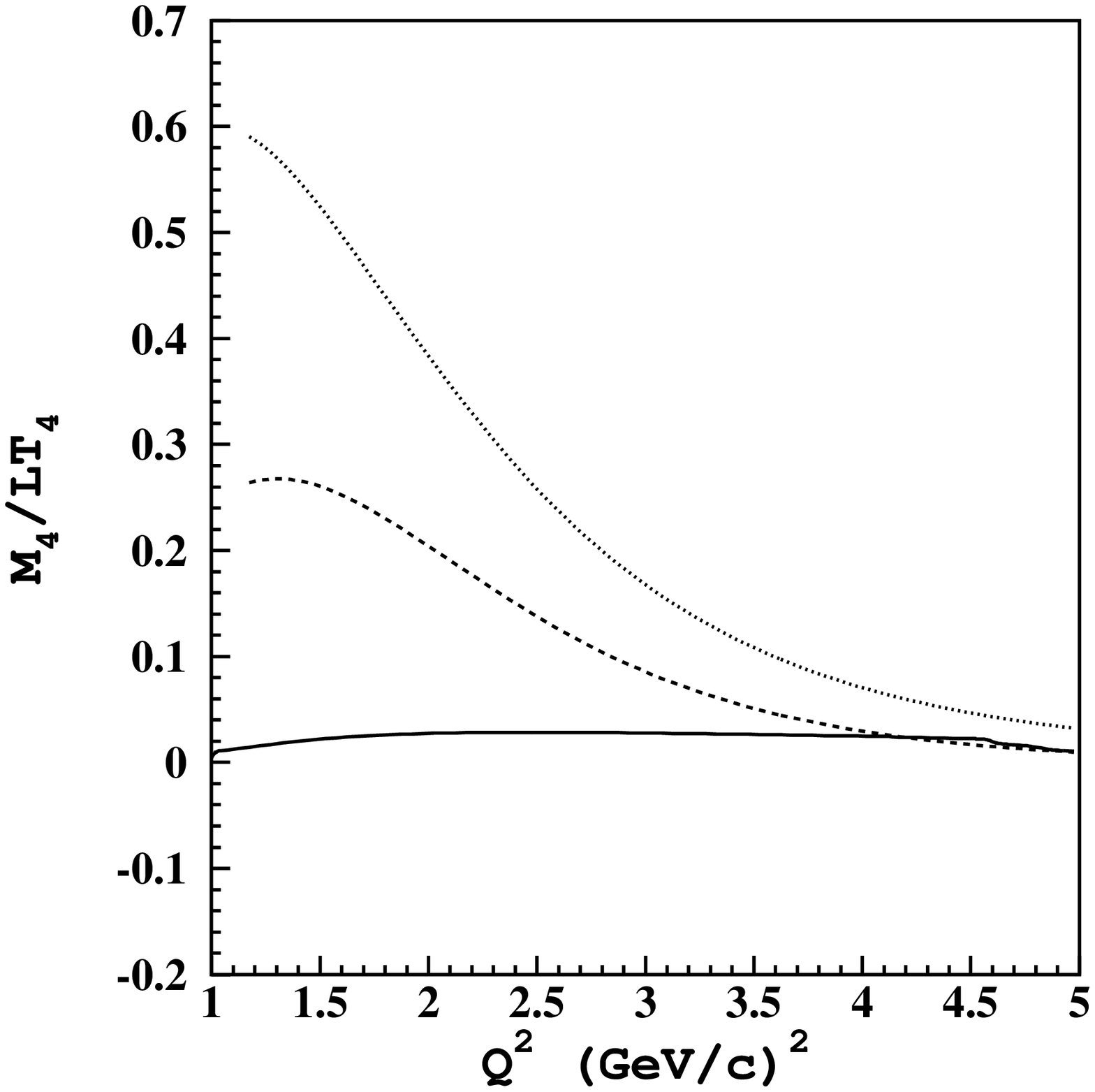}
\includegraphics[bb=1cm 4cm 20cm 24cm, scale=0.4]{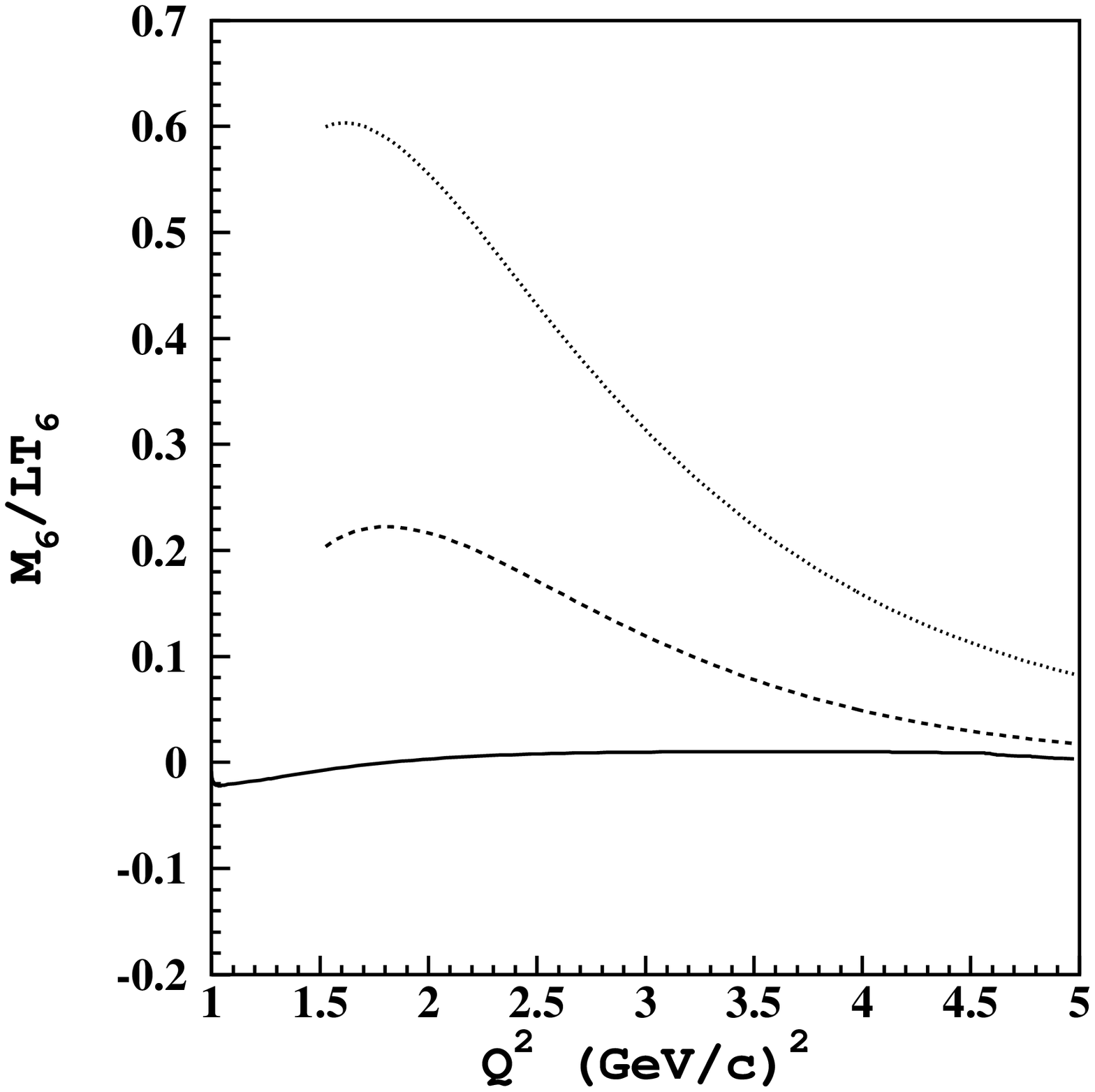}
\includegraphics[bb=1cm 4cm 20cm 24cm, scale=0.4]{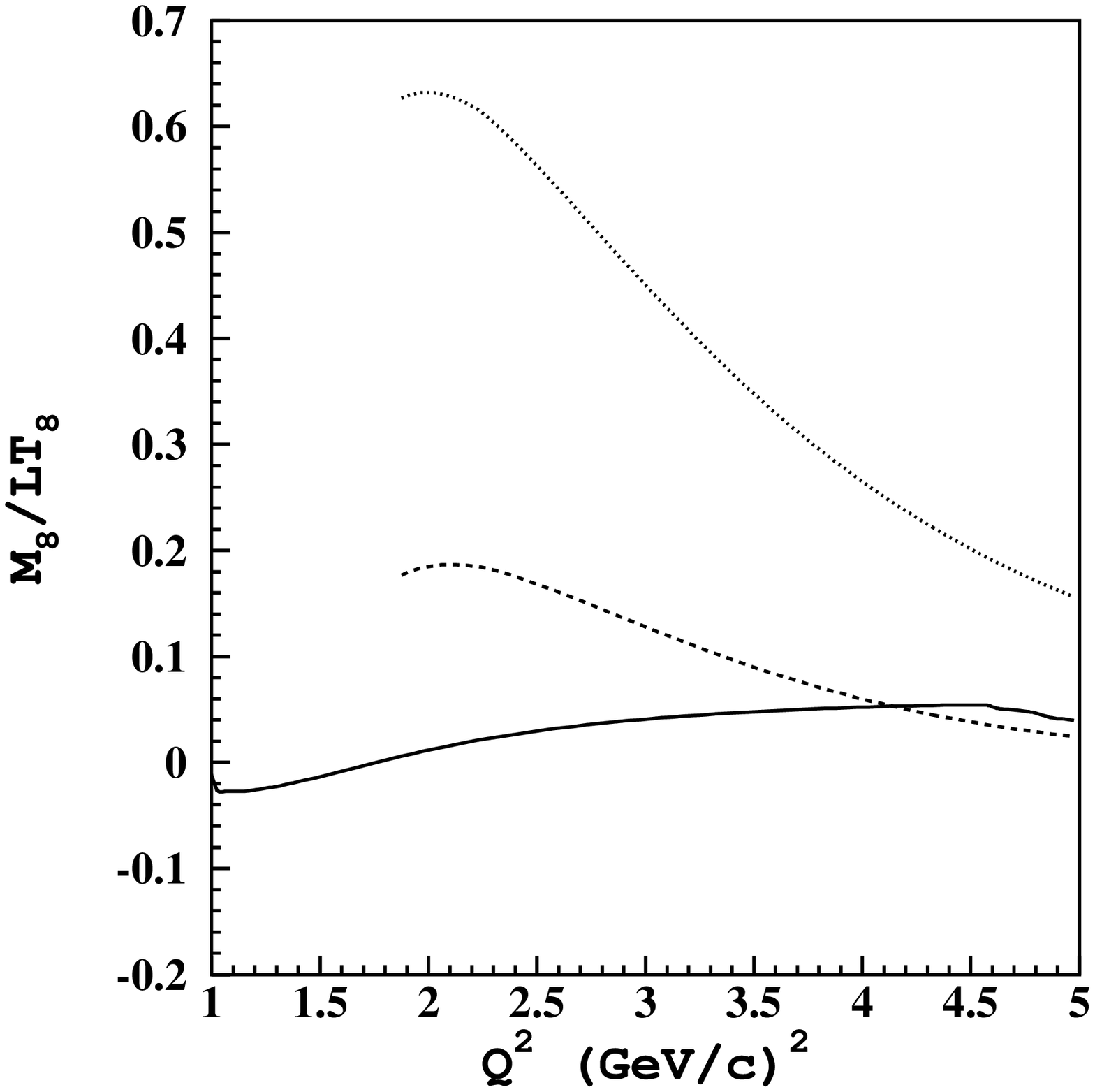}
\caption{\label{fig:rmin} The total higher twist contribution and
the elastic+resonance contribution to the $F_2^p$ structure function moments
relative to its leading twist part taken from Ref.~\cite{Osipenko_f2p}:
solid line - the total higher twist contribution;
dashed line - the resonance contribution using resonance
form-factors from Ref.~\cite{Simula_ffs};
dotted line - the elastic+resonance contribution using resonance
form-factors from Ref.~\cite{Simula_ffs}.}
}

\section{Conclusions}
We calculated the contribution of the nucleon elastic pole and
its excited states into
moments of the nucleon structure function $F_2$.
Due to formation of an intermediate bound state this contribution
cannot be described by an incoherent elastic scattering
of the virtual photon off a quark (either constituent or current).
In terms of OPE this contribution belongs to higher twists.
This is confirmed by the observed $Q^2$-dependence of estimated
the resonance contribution shown in the Fig.~\ref{fig:rmin}.
Vice versa, assuming that conceptually the resonance contribution
as well as the elastic peak contribution
should belong to higher twists one obtains a stringent
test on the large-$Q^2$ behavior of resonance form-factors.

We found that the elastic peak contribution is actually twice smaller
than it was known before. The reduction of the elastic contribution
in moments leads to an improvement of the duality effect in the low $Q^2$
region. As you can see in Fig.~\ref{fig:mdual} the second moment becomes
almost constant while higher moments are much more flat after the correction
and leave no room for a significant higher twist contribution (see also Fig.~\ref{fig:rmin}).
This leads to the conclusion that the extraction of nucleon elastic form-factors
from DIS pioneered in Ref.~\cite{Rujula} is not hopeless and the large
discrepancies found more recently in Ref.~\cite{Simula-el} are due to
missing 0.5 factor.

In Refs.~\cite{Osipenko_f2p,Osipenko_f2d} we have seen that different higher twist terms
cancel generating the well known duality phenomenon. The resonance
contribution estimated in this article yields coherent positive
higher twist in the moments. Additional positive contribution
to higher twists is given by the elastic scattering.
While the negative higher twist which largely cancels the resonance
and elastic contributions is related to number of exclusive channel
thresholds (e.g. single pion production threshold). If all elastic, resonance
and the threshold contributions are subtracted from the moments
the remaining part should be related to the incoherent
scattering off proton constituents regardless $Q^2$ value.
Therefore, an estimate of the threshold contribution to the structure
function moments would allow to study $Q^2$-behavior of
form-factors of proton constituents.

Notice that both elastic and resonance higher twist contributions
are related to nucleon and resonance bound state wave functions.
Therefore these terms cannot be described by asymptotic freedom methods
like Borel summation of the leading twist $\alpha_S$ series which
also generates $1/Q^2$ corrections. Given large magnitude of elastic and
resonance contributions in the total moments (in particular for $n>2$)
the good overlap between renormalon-based
fits~\cite{Gardi_htw} and the data should be taken as accidental or related
to the duality phenomenon.

It is clear from Fig.~\ref{fig:rmin} that the resonance contribution
itself does not follow DIS scaling behavior. Therefore, the observed
approximate scaling of integrated resonance peaks (known as local duality)
is due to compensation between rising resonances and decreasing background
at low $Q^2$.

Large physics program focused on extraction of nucleon resonance
form-factors from electro- and photo-production data undergoing
in Jefferson Lab will allow for precise evaluation of the main higher
twist contribution to the nucleon structure functions.
\begin{acknowledgments}
We are grateful to S. Simula for fruitful discussions.
\end{acknowledgments}



\begin{thebibliography}{999}

\bibitem{SGR} S. Simula,
   {\it Phys. Lett.}
   {\bf B493} (2000) 325

\bibitem{Osipenko_f2p} M. Osipenko {\it at al.},
   {\it Phys. Rev.}
   {\bf D67} (2003) 092001

\bibitem{Osipenko_f2d} M. Osipenko {\it at al.},
   {\it Phys. Rev.}
   {\bf C73} (2006) 045205

\bibitem{Osipenko_f2n} M. Osipenko {\it at al.},
   {\it Nucl. Phys.}
   {\bf A766} (2006) 142

\bibitem{Ricco} G. Ricco {\it at al.},
   {\it Nucl. Phys.}
   {\bf B555} (1999) 306

\bibitem{Ricco2} G. Ricco {\it at al.},
   {\it Phys. Rev.}
   {\bf C57} (1998) 356

\bibitem{Rujula} A. De Rujula, H. Georgi and H. Politzer,
   {\it Ann. Phys.}
   {\bf 103} (1977) 315

\bibitem{Ji} X. Ji and P. Unrau,
   {\it Phys. Rev.}
   {\bf D52} (1995) 72

\bibitem{Bodek} U. Yang and A. Bodek,
   {\it Eur. Phys. J.}
   {\bf C13} (2000) 241

\bibitem{Nikulesku} C. S. Armstrong {\it at al.},
   {\it Phys. Rev.}
   {\bf D63} (2001) 094008

\bibitem{Shurak} E. V. Shuryak and A. I. Vainshtein,
   {\it Nucl. Phys.}
   {\bf B201} (1982) 141

\bibitem{BloGil} E. Bloom and F. Gilman,
   {\it Phys. Rev. Lett.}
   {\bf 25} (1970) 1140

\bibitem{Isgur} N. Isgur {\it at al.},
   {\it Phys. Rev.}
   {\bf D64} (2001) 054005

\bibitem{Petronzio} R. Petronzio, S. Simula and G. Ricco,
   {\it Phys. Rev.}
   {\bf D67} (2003) 094004

\bibitem{Roberts} R. G. Roberts,
   {\it The structure of the proton},
   Cambridge University Press 1990.

\bibitem{Weise} J. Edelmann {\it at al.},
   {\it Nucl. Phys.}
   {\bf A665} (1999) 125

\bibitem{Thomas} A. W. Thomas and W. Weise,
   {\it The structure of the nucleon},
   WILEY-VCH 2001.

\bibitem{Drehsel_oth} D. Drechsel {\it at al.},
   {\it Phys. Rep.}
   {\bf 378} (2003) 99

\bibitem{SQTM} V. D. Burkert {\it at al.},
   {\it Phys. Rev.}
   {\bf C67} (2003) 035204

\bibitem{Simula_ffs} S. Simula {\it at al.},
   {\it Phys. Rev.}
   {\bf D65} (2002) 034017

\bibitem{PDG} D. E. Groom {\it at al.},
   {\it Eur. Phys. Jour.}
   {\bf C15} (2000) 1

\bibitem{Melnitc_tmc} F. M. Stefens and W. Melnitchouk,
   {\it Phys.\ Rev.}
   {\bf C73} (2006) 055202

\bibitem{Simula-el} S. Simula,
   {\it Phys. Lett.}
   {\bf B481} (2000) 14

\bibitem{Gardi_htw} E. Gardi and R.G. Roberts,
   {\it Nucl. Phys.}
   {\bf B653} (2003) 227

\end{thebibliography}
\end{document}